\documentclass[conference]{IEEEtran}
\usepackage{booktabs}
\usepackage{cite}
\usepackage{amsmath,amssymb,amsfonts}
\usepackage{algorithmic}
\usepackage{graphicx}
\usepackage{textcomp}
\usepackage{xcolor}
\usepackage{tabularx}
\usepackage{makecell}  
\usepackage{url}

\def\BibTeX{{\rm B\kern-.05em{\sc i\kern-.025em b}\kern-.08em
    T\kern-.1667em\lower.7ex\hbox{E}\kern-.125emX}}
\begin{document}

\title{A Temporal-Envelope Frontend with Learnable Per-Channel Energy Normalization for Whisper-Based Children's ASR
}


\author{\IEEEauthorblockN{Edem Ahadzi, Ruchi Pandey, Tomi H. Kinnunen}
\IEEEauthorblockA{\textit{School of Computing} \\
\textit{University of Eastern Finland}\\
Joensuu, Finland \\
edem.ahadzi@uef.fi, ruchi.pandey@uef.fi, tomi.kinnunen@uef.fi}
}
\maketitle

\begin{abstract}
Temporal envelopes carry cues critical to speech intelligibility, yet ASR frontends based on log-mel spectrograms do not explicitly model continuous sub-band envelope structure. This limitation is particularly acute for children's speech, where high acoustic variability demands robust feature representations. We propose a modular time-domain frontend that decomposes speech into sub-band envelopes using mel-spaced windowed-sinc filters and the Hilbert transform, with learnable per-channel energy normalization (PCEN) jointly optimized with the Whisper model. On the MyST children's speech corpus, systematic ablations identify full-band windowed-sinc filters, Hilbert envelopes, a 25\,Hz smoothing cutoff, and learnable PCEN as the best configuration. Under the same Whisper-small fine-tuning setup, the frontend reduces WER from $13.16\%$ to $11.08\%$, a $15.8\%$ relative reduction over the log-mel baseline, and outperforms the evaluated Kid-Whisper checkpoint on the same cleaned test split. These results show that temporal-envelope representations and learnable frontend normalization are effective complements to backend adaptation for children's ASR.
\end{abstract}


\section{Introduction}

Children's automatic speech recognition (ASR) remains challenging due to acoustic variability and scarcity of children's speech data~\cite{attia2024kidwhisperbridgingperformancegap, Ahadzi_2025, GELIN202171}. Compared to adults, children have smaller vocal folds and shorter vocal tracts, resulting in higher fundamental (F0) and formant frequencies respectively, as well as greater variability in F0, articulation, and speaking rate~\cite{children9111690, improving_childrens_speech, singh2025causalstructurediscoveryerror}. Children's speech also exhibits irregular breath control, sudden loudness fluctuations, and imprecise articulation~\cite{liu2022investigationapplyingacousticfeature}. As a result, ASR systems perform notably worse on children's speech under otherwise comparable conditions~\cite{jain2023adaptationwhispermodelschild, jain2023sslmethodsimprovechildrenspeech}.

Prior work has addressed this mismatch primarily through model-level interventions. These include fine-tuning foundational models on children's data~\cite{jain2023adaptationwhispermodelschild,attia2024kidwhisperbridgingperformancegap}, vocal tract length perturbation and F0-based augmentation~\cite{kathania2022spectralwarping,fan2024benchmarkingchildrensasrsupervised, singh2024childaugmentdataaugmentationmethods}, and parameter-efficient transfer learning \cite{liu2024sparselysharedlorawhisper}, to name a few. However, all of these approaches retain the mel-spectrogram frontend used in systems like Whisper~\cite{radford2022robustspeechrecognitionlargescale}. In this work, we focus on this critically important \emph{acoustic front-end}, a component that has received surprisingly little attention in children's ASR compared to model-level interventions.

Although designed to approximate human auditory perception~\cite{stevens1937scale, volkmann2005mel}, the commonly-used mel-spectrogram representation may not well capture the distinct spectral and temporal characteristics of children's speech. Its \emph{frame-based} spectral representation extracts features independently from local parts of the waveform and does not explicitly preserve the \emph{continuous} temporal envelope structure within frequency bands, known to be important for speech intelligibility~\cite{rosen1992temporal, shannon1995speech}. \looseness=-1
\begin{figure*}[t!]
  \centering
  \includegraphics[width=0.93\linewidth]{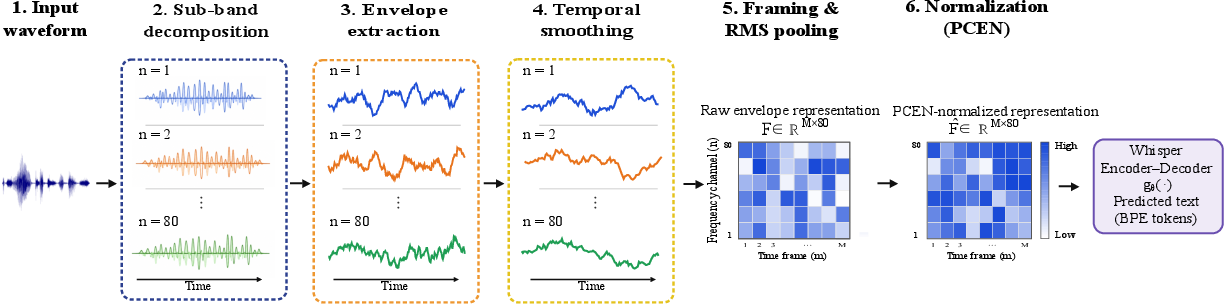}
  \caption{Overview of the proposed temporal-envelope frontend. The waveform is decomposed into K=80 mel-spaced sub-bands via Hamming-windowed sinc filters (1--2), instantaneous envelopes are extracted using the Hilbert transform (3), smoothed with a 25\,Hz zero-phase low-pass filter (4), segmented into 25\,ms frames and RMS-pooled at a 100\,Hz frame rate (5), and normalized with learnable per-channel PCEN (6) before being passed to the Whisper encoder--decoder.}
  \label{fig:frontend_pipeline}
  \vspace{-0.5cm}
\end{figure*}

Indeed, prior psychoacoustic research has demonstrated that human listeners can achieve near-perfect speech recognition from temporal envelopes extracted from as few as three frequency bands~\cite{shannon1995speech}. This indicates that amplitude modulation patterns alone can carry enough intelligibility information. Unlike log-mel spectrograms, which represent spectral energy at each time frame, envelope-based features explicitly capture the energy fluctuation over time within frequency bands. These envelopes encode syllabic rhythm ($\sim$4\,Hz) and phonemic information ($\sim$15--50\,Hz), while preserving rapid spectrotemporal changes associated with consonant bursts and formant transitions \cite{rosen1992temporal, ding2017temporal}. These acoustic cues are especially variable in children's speech due to their developing articulatory motor control \cite{lee1999acoustics, gerosa2007acoustic}. 
This motivates the exploration of alternative frontends better aligned with children's speech.

As such, temporal-domain speech analysis, including feature extraction from sub-band temporal-envelope features, is not new to audio processing~\cite{zeghidour2021leaflearnablefrontendaudio, sadjadi2015mhec} or ASR tasks. Prior to deep learning based ASR, sub-band envelope features were used for noise-robust \emph{adult} ASR, e.g.\ full-wave rectified envelopes fed to a time-delay neural network (TDNN) denoising autoencoder~\cite{do2017improved}, or an amplitude-modulation filterbank augmenting conventional Mel-frequency cepstral coefficient (MFCC) with Gaussian mixture model (GMM) and deep neural network (DNN) back-ends~\cite{moritz2015auditory}. More recently, sub-band envelopes over linearly-spaced non-overlapping filters have been utilized in children's keyword spotting in a zero-resource setting, with a fixed log-compression front-end and a DNN-hidden Markov model (HMM) back-end~\cite{DAS2026101954}. Most closely related, \cite{ANKITA2024104385} estimate sub-band envelopes using frequency domain linear prediction (FDLP) features to train an end-to-end children's ASR model from scratch. We instead estimate envelopes directly in the time domain and use them to replace the log-mel input of a pre-trained foundation model. \textbf{To our knowledge, however, temporal-envelope frontends have not been used to replace the input representation of a pre-trained end-to-end speech foundation model for children's ASR.} The modeling of long-range temporal context is typically deferred to the back-end, where self-attention aggregates context across frames~\cite{vaswani2023attentionneed, radford2022robustspeechrecognitionlargescale}. However powerful, these mechanisms cannot recover what the acoustic frontend discards.  As our experiments will demonstrate, replacing the commonplace mel-spectrogram front-end used in Whisper~\cite{radford2022robustspeechrecognitionlargescale} with an alternative temporal envelope processing motivated approach translates to practical improvements in children's ASR. Thus, our proposed methodology is intended to \emph{complement} the already-powerful deep learning ASR.

Unlike prior envelope-based work in ASR, which relied on non-learnable statistical normalizers, our normalization layer is learnable and jointly optimized end-to-end with the ASR back-end.
To the best of our knowledge, this is the first systematic investigation of temporal envelope frontends for children's ASR within modern deep-learning-based systems. We conduct a comprehensive ablation on MyST, a standard children's ASR benchmark, isolating each frontend stage to find the configuration best suited to children's speech. 
\footnote{\raggedright Code at \url{https://github.com/ahadziii/temporal-envelope-whisper-asr} and flagged-utterance list available on request.}


\section{Standard Whisper Mel-Spectrogram Frontend}
\label{sec:conv_frontend}
The Whisper frontend \cite{radford2022robustspeechrecognitionlargescale}, adopted as our baseline, 
uses mel-spectrogram features. The waveform is segmented into 25\,ms Hann-windowed \cite{oppenheim1999discrete} frames with a 10\,ms hop, followed by a discrete Fourier transform (DFT). The squared magnitude of the resulting complex-valued spectrum yields the power at each frequency bin, projected onto 80 mel-spaced triangular filters \cite{volkmann2005mel} to reduce the dimensionality. Finally, logarithmic compression is applied to the mel-filter energies to approximate human loudness perception and to reduce dynamic range. Whisper implements this as \emph{clipped log-affine rescaling} (detailed in Section~\ref{sec:normalization}), producing the feature matrix $\mathbf{F} \in \mathbb{R}^{M \times K}$ where $K\!=\!80$ is the number of frequency channels and $M\!=\!3000$ is the number of frames, corresponding to Whisper's fixed 30-second input window (shorter utterances are zero-padded) at a 100\,Hz frame rate. For a given mel bin $n$ and frame $m$, the mel-spectrogram pipeline is expressed as: \looseness=-1
\begin{equation}
F_n(m) = \log \left( \sum_{k} H_n(k) \left| \mathrm{DFT}\{x_m\}(k) \right|^2 \right),
\label{eq:conv_pipeline}
\end{equation}
where $x_m$ is the $m$-th windowed frame and $H_n(k)$ denotes the triangular mel filter weights for the $n$-th channel, with center frequencies equispaced on the mel scale \cite{volkmann2005mel} defined as $\mathrm{mel}(f) = 2595 \log_{10}(1 + f/700)$.

\section{Proposed Adaptive Frontend}
\label{sec:frontend}

We replace Whisper's mel-spectrogram frontend (Section~\ref{sec:conv_frontend}) with a learnable, time-domain alternative. It follows an analogous sequence of frequency decomposition, energy extraction, and dynamic-range compression, differing in two respects. Our proposed frontend characterises each sub-band through its temporal envelope rather than a short-term spectral snapshot, and its normalization is optimized jointly with the back-end rather than hand-crafted. As illustrated in Figure~\ref{fig:frontend_pipeline}, the frontend applies time-domain bandpass filters to decompose the waveform into narrowband sub-band signals. A non-linear envelope extraction step~\cite{assaneo2024speech} recovers the instantaneous amplitude within each sub-band. The envelopes are then smoothed to isolate modulation patterns relevant to intelligibility~\cite{rosen1992temporal, ding2017temporal}, before being framed and pooled to yield a non-negative energy estimate per frame. A final normalization step compensates for acoustic variability across speakers and recording conditions. The following subsections detail the candidate filterbanks, envelope extractors, and normalization schemes we consider. Each stage offers alternatives we evaluate empirically and the configuration selected by our ablation (Section~\ref{sec:results}) constitutes the proposed system. Throughout, we mark each alternative as the proposed choice or a baseline-under-test.

\subsection{Filtering}
The first stage decomposes the speech signal, $\mathbf{x}$ into $K$ frequency sub-bands by convolving it with a bank of time-domain band-pass filters $(\phi_n)_{n=0}^{K-1}$. Unlike~\eqref{eq:conv_pipeline} which applies 
mel-filters to the short-term power spectrum, our filters operate directly in the time domain to isolate sub-band signals, preserving temporal fine-structure. 
We consider two alternative time-domain filterbanks, both with center frequencies fixed on the mel scale. Of these, the mel-spaced windowed-sinc filterbank is our proposed choice; the Gabor filterbank serves as a baseline-under-test (Section~\ref{sec:results}).

\subsubsection{\textbf{Mel-Spaced Windowed-Sinc Filterbank}}
In this parameterization, each bandpass filter is a finite impulse response (FIR) Hamming-windowed sinc kernel \cite{oppenheim1999discrete}, $\ell_2$-normalized, with center frequencies $\{f_0, f_1, \dots, f_{K-1}\}$ equispaced on the mel scale. For $n = 1, \dots, K{-}2$, the passband is defined by adjacent center frequencies
$f_{\mathrm{low}}^{(n)}\!=\!f_{n-1}$ and $f_{\mathrm{high}}^{(n)}\!= \!f_{n+1}$, producing overlapping passbands with approximately flat magnitude response. The sub-band signals are obtained using 
\begin{equation}
x_n[t] = (\mathbf{x} *_{\mathrm{zp}} \hat{\phi}_n)[t], \qquad n = 0, \ldots, K-1
\label{eq:fir_bp}
\end{equation}
where $\hat{\phi}_n \in \mathbb{R}^L$ is the normalized bandpass kernel and $*_{\mathrm{zp}}$ denotes zero-phase filtering via symmetric 
padding of $\lfloor L/2 \rfloor$ samples at both ends, eliminating phase distortion and preserving temporal alignment across sub-bands.

\subsubsection{\textbf{Gabor Filterbank}}
While the windowed-sinc kernel is a common filter design choice, it does not optimise the time-frequency resolution trade-off. As an alternative, we consider Gabor filters~\cite{gabor1946theory} which, as Gaussian-modulated sinusoids, achieve the theoretical lower bound of the time-frequency trade-off~\cite{mallat2009wavelet}. The $n$-th filter is a Gaussian-modulated cosine:
\begin{equation}
g_n(t) = e^{-t^2/(2\sigma_n^2)} \cos(2\pi f_n t),
\label{eq:gabor}
\end{equation}
where $f_n$ is the center frequency and $\sigma_n$ controls the temporal extent of the Gaussian window. The Gaussian width ($\sigma_n$) is set via the $-3$\,dB bandwidth relationship $\sigma_n = \sqrt{2 \ln 2}\,/\,(\pi \cdot B_n)$, where $B_n = (f_{n+1} - f_{n-1})/2$ is the average spacing between two adjacent center frequencies. Each Gabor kernel is $\ell_2$-normalized, and sub-band signals are obtained via the same zero-phase procedure as in Equation~\eqref{eq:fir_bp}. The Gabor kernel yields a Gaussian spectral profile rather than the approximately flat passband of the windowed-sinc design.

\subsection{Envelope Extraction}
\label{sec:envelope}
The second stage extracts the instantaneous temporal envelope from each sub-band signal $x_n(t)$, computed via either filtering method above. We consider two widely used envelope extractors: Hilbert transform, our proposed choice, and the Square Modulus as a simpler baseline.

\subsubsection{\textbf{Hilbert transform}}
The Hilbert transform $\mathcal{H}\{\cdot\}$ applies a $90^{\circ}$ phase shift to all 
frequency components. 
Applying it to $x_n(t)$ yields the analytic signal $z_n(t) = x_n(t) + j\,\mathcal{H}\{x_n(t)\}$, a complex-valued representation that retains only positive frequency components, whose modulus gives the instantaneous envelope \cite{oppenheim1999discrete, sadjadi2015mhec}:
\begin{equation}
E_n^{\mathrm{H}}(t) = |z_n(t)| = \sqrt{x_n(t)^2 + \mathcal{H}\{x_n(t)\}^2}.
\label{eq:hilbert_env}
\end{equation}

\subsubsection{\textbf{Square Modulus}}
As a simpler alternative, the squared modulus approximates the envelope as instantaneous power $E_n^{\mathrm{S}}(t) = |x_n(t)|^2$~\cite{zeghidour2021leaflearnablefrontendaudio}, bypassing analytic signal construction.

\subsection{Temporal Smoothing and Framing}
\label{sec:smoothing}
The output of the envelope extraction stage retains fast oscillations near the sub-band carrier frequency. To isolate the slowly varying temporal envelope, relevant to syllabic and phonemic cues~\cite{rosen1992temporal,ding2017temporal}, we smooth each envelope with a zero-phase low-pass FIR filter $\tilde{E}_n(t) = (E_n *_{\mathrm{zp}} h_{\mathrm{LP}})(t)$,
where $\tilde{E}_n(t)$ is the smoothed envelope and $h_{\mathrm{LP}}(t)$ is a low-pass FIR filter with cutoff frequency $f_c$. The smoothed envelopes are then segmented into 25\,ms frames at a 10\,ms hop and pooled via root-mean-square (RMS) computation, yielding the feature matrix $\mathbf{F} \in \mathbb{R}^{M \times K}$ at a 100\,Hz frame rate. 

\subsection{Normalization}
\label{sec:normalization}
The final stage normalizes the time-frequency representation $\mathbf{F}$ to reduce dynamic range and mitigate acoustic variability across speakers and recording conditions. Our proposed frontend uses the learnable PCEN scheme described below; the remaining strategies 
are non-learnable baselines against which we test it.

\subsubsection{\textbf{Clipped Log-affine Rescaling}}
As a baseline, we adopt Whisper's~\cite{radford2022robustspeechrecognitionlargescale} default normalization.
In Whisper \cite{radford2022robustspeechrecognitionlargescale}, the sub-band features are floored at $10^{-10}$ to avoid numerical instability, log-compressed, clipped to within 80\,dB (i.e., a range of 8.0 in $\log_{10}$ scale) of the utterance-level peak and mapped to an approximate range of $[-1, 1]$, using
\begin{equation}
\hat{F}_n(m) =
\frac{
\max(\log_{10}(F_n(m)), L_{\max} - 8.0) + 4.0
}{4.0}
\label{eq:whisper_norm},
\end{equation}
where $ L_{\max} = \max_{n,m} \log_{10}(F_n(m))$ is the utterance-level peak log-magnitude.

\subsubsection{\textbf{Mean (MN) and Mean--Variance (MVN) normalization}}
We consider classic normalization techniques as non-learnable baselines for suppressing linear channel or speaker related effects. These methods originate in the cepstral domain, where they are conventionally denoted as cepstral mean (CMN)~\cite{atal1974effectivenessofLP, furui1981cepstralanalysis} and mean--variance (CMVN)~\cite{viikki1998cepstral} normalization. Since our features are sub-band envelope energies rather than cepstral coefficients, we apply the same mean and variance normalization directly to those energies and refer to them here as MN and MVN. MN, a cornerstone of early ASR and speaker verification systems, subtracts the per-channel mean, while its variance-normalized extension, MVN, additionally scales by the standard deviation:
\begin{equation}
\hat{F}_n(m) = \frac{F_n(m) - \mu_n}{s_n}
\label{eq:cmn_cmvn}
\end{equation}
where $\mu_n$ is the per-channel sample mean and $s_n$ is a scale factor: $s_n = 1$ for MN and $s_n = \sigma_n$ (the sample standard deviation) for MVN. Under a convolutive noise model, mean subtraction removes constant channel or speaker-dependent bias~\cite{kinnunen2010overview}, while variance normalization standardizes the energy distribution across channels. For both methods, we evaluate utterance-level and channel-level variants, where statistics are computed across all channels jointly or per channel independently. We additionally consider a sliding-window variant of MVN, where statistics are computed over a local window of 300 frames (corresponding to a 3\,s temporal context at a 100\,Hz frame rate) centered at each time step. This allows the normalization to adapt to time-varying conditions within an utterance, rather than relying on global utterance-level statistics. \looseness=-1

\subsubsection{\textbf{Per-Channel Energy Normalization (PCEN)}}
PCEN~\cite{wang2016trainablefrontendrobustfarfield} is our proposed normalization, introducing four parameters $(s, \alpha, \delta, r)$ jointly optimized with the ASR backend. While MN and MVN adapt to each utterance through fixed statistical estimation, PCEN provides a learnable pipeline of temporal smoothing, adaptive gain control (AGC), and dynamic range compression \cite{Pcen_how_why}. The sub-band energies $E(t,f) \equiv F_n(m)$ are first smoothed with a first-order IIR filter, $M(t,f) = (1 - s)\,M(t{-}1,f) + s\,E(t,f)$, where $s \in (0,1]$ controls the smoothing time constant, to estimate the local energy context. An AGC stage then normalizes loud and quiet regions relative to their local context. Lastly, a compressive nonlinearity reduces the remaining dynamic range:\looseness=-1
\begin{equation}
\mathrm{PCEN}(t,f) = \left(\frac{E(t,f)}{(\varepsilon + M(t,f))^{\alpha}} + \delta\right)^{r} - \delta^{r},
\label{eq:pcen}
\end{equation}
where $\alpha \in [0,1]$ controls the AGC strength, $\delta > 0$ is a bias offset stabilising compression for near-silent frames, $r \in (0,1]$ is the compression exponent, and $\varepsilon$ is a small constant for numerical stability.

\section{Experimental Setup}
\label{sec:experimentaldesign}

\subsection{Dataset}
\label{subsec:dataset}
We conduct our experiments on the My Science Tutor (MyST) children's conversational speech corpus~\cite{pradhan2023sciencetutormyst}. MyST contains roughly 197 hours of transcribed conversational speech, recorded from virtual tutoring sessions in physics, geography, biology, and other topics from 1371 students in grades 3--5. While MyST is widely adopted for children's ASR research and comes with a standard evaluation protocol, recent work has identified a number of problematic recordings that can bias evaluation~\cite{attia2024kidwhisperbridgingperformancegap}. This includes utterances with transcription errors, audio-transcript misalignment, and degraded audio quality. To this end, and in order to ensure our reported results are comparable with prior work, we follow identical data-cleaning pipeline as reported in~\cite{attia2024kidwhisperbridgingperformancegap}. This includes removing utterances flagged as low quality using a refined flagged-utterance list\footnote{This list was obtained through personal communication with the authors of~\cite{attia2024kidwhisperbridgingperformancegap}.}, discarding utterances with fewer than three words since they lack the context to distinguish homophones, and excluding utterances longer than 30\,s, which exceed Whisper's input limit. The original MyST train, development, and test splits are retained to avoid speaker overlap across partitions. After applying the above filtering steps, the resulting dataset comprises 136.27 hours of training data, 21.49 hours of development data and 23.26 hours of test data, corresponding to approximately 91.9\% of the original transcribed data.

\subsection{Model and Training Configuration}
We adopt OpenAI's \emph{Whisper-small}~\cite{radford2022robustspeechrecognitionlargescale}, a transformer-based encoder--decoder~\cite{vaswani2023attentionneed} trained on 680,000 hours of weakly supervised audio, with its default log-mel frontend (Section~\ref{sec:conv_frontend}) as our baseline. Against this baseline, we evaluate our proposed frontend through a modular ablation study, progressively optimizing each pipeline stage: \textbf{(1)} envelope extraction method (Hilbert vs.\ squared modulus) and low-pass cutoff $f_c \in \{15, 25, 50\}$\,Hz, using Whisper's clipped log-affine rescaling as default normalization; \textbf{(2)} normalization strategy (MN, MVN, and sliding-window MVN, each at utterance and channel level); \textbf{(3)} filterbank architecture (windowed-sinc vs.\ Gabor, at 0--8\,kHz and 150--8\,kHz, where the 150\,Hz lower bound was determined via pitch tracking with Parselmouth~\cite{jadoul2018parselmouth}); and \textbf{(4)} replacement of non-learnable normalization with learnable PCEN, initialised with per-channel parameters $s = 0.025$, $\alpha = 0.98$, $\delta = 2.0$, $r = 0.5$, $\varepsilon = 10^{-6}$ following~\cite{wang2016trainablefrontendrobustfarfield}. 
\begin{figure}[t]
  \centering
  \includegraphics[width=0.95\linewidth]{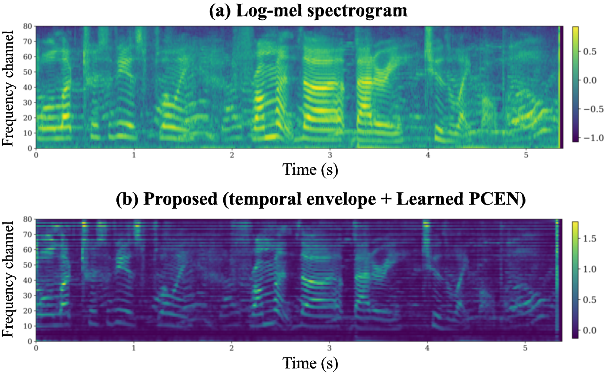}
  \caption{Feature representations for the utterance \textit{``well electricity is flowing from the battery to the light bulb''}, spoken by a child.}
  \label{fig:spectrogram}
  \vspace{-0.35cm}
\end{figure}
Let $\mathbf{x}_i \in \mathbb{R}^{T_i}$ denote the $i$-th input waveform of $T_i$ samples and $\mathbf{y}_i = (y_{i,1}, \dots, y_{i,L_i})$ the corresponding target transcription $L_i$ tokens, where $i \in \{1, \dots, N\}$ indexes the training utterances. Our frontend $\mathcal{F}_\psi$, parametrized by $\psi$, maps the waveform to a time-frequency representation $\mathcal{F}_\psi(\mathbf{x}_i) \in \mathbb{R}^{M_i \times K}$ with $M_i$ time frames and $K$ frequency channels. The resulting representation is processed by the Whisper encoder--decoder $g_\theta$, parameterized by $\theta$. The frontend and backend are trained jointly by minimising the cross-entropy loss over the training set:\looseness=-1
\begin{equation}
(\theta^*, \psi^*) =
\arg\min_{\theta, \psi}
\frac{1}{N} \sum_{i=1}^{N}
\sum_{t=1}^{L_i}
\mathcal{L}
\bigl(g_\theta(\mathcal{F}_\psi(\mathbf{x}_i))_t, y_{i,t}\bigr),
\label{eq:asr_objective}
\end{equation}
where $\mathcal{L}$ is the standard cross-entropy loss at each decoding step 
used in sequence-to-sequence ASR~\cite{radford2022robustspeechrecognitionlargescale}, $g_\theta(\mathcal{F}_\psi(\mathbf{x}_i))_t$ is the predicted distribution over the vocabulary at position $t$ for utterance $i$, and $y_{i,t}$ is the ground-truth token from Whisper's byte pair encoding (BPE) vocabulary. Since the encoder-decoder uses cross-attention, no explicit alignment between the $M_i$ input frames and $L_i$ output tokens is required~\cite{vaswani2023attentionneed}. \looseness=-1

For all configurations, the Whisper backend was fine-tuned for two epochs at a learning rate (LR) of $1.75 \times 10^{-5}$, chosen to preserve the pre-trained representations and mitigate catastrophic forgetting~\cite{kirkpatrick2017overcoming, Ahadzi_2025}. In the PCEN configurations, the frontend parameters form a separate optimiser group with a learning rate of $1 \times 10^{-2}$. Only the four PCEN parameters $(s,\alpha,\delta,r)$, learned independently per channel, are trainable while the filterbank remains fixed at its mel-scale initialization. These PCEN parameters are updated jointly with the backend by backpropagating the cross-entropy loss of Equation~\eqref{eq:asr_objective}.

\section{Results and Discussions}
\label{sec:results}
Tables~\ref{tab:envelope}--\ref{tab:filterbank_pcen} summarise results across the varying frontend configurations. The baseline Whisper model with its standard log-mel spectrogram frontend achieves a WER of 13.16\% on the MyST test set, serving as reference for all subsequent comparisons. Across these configurations, the backend fine-tuning is identical, so the frontend is the only varied factor and the reported gains isolate its contribution. 

\begin{table}[t]
\caption{Effect of envelope extraction method and low-pass cutoff frequency.
Filterbank: Mel-Spaced Windowed-Sinc (0--8\,kHz); normalization: clipped log-affine rescaling.}
\label{tab:envelope}
\centering
\setlength{\tabcolsep}{2pt}
\renewcommand{\arraystretch}{0.9}
\footnotesize
\begin{tabular}{@{}lccc@{}}
\toprule
& \multicolumn{3}{c}{WER (\%)} \\
\cmidrule(l){2-4}
Whisper baseline (log-mel) & \multicolumn{3}{c}{13.16} \\
\midrule
Envelope & $f_c = 15$\,Hz & $f_c = 25$\,Hz & $f_c = 50$\,Hz \\
\midrule
Hilbert          & 12.20& \textbf{11.99}& 12.96\\
Squared mod.     & 12.24& 12.06& 13.03\\
\bottomrule
\end{tabular}
\end{table}

As shown in Table~\ref{tab:envelope}, the Hilbert transform outperformed the squared modulus across all cutoff frequencies, achieving a WER of 11.99\% at $f_c{=}25$\,Hz. Moving the cutoff in either direction degraded performance, identifying 25\,Hz as the best balance between temporal detail and robustness.


\begin{table}[t]
\caption{Effect of normalization strategy. With Mel-Spaced Windowed-Sinc filters (0--8\,kHz), Hilbert envelope, $f_c{=}25$\,Hz.}
\label{tab:norm}
\centering
\setlength{\tabcolsep}{4pt}
\renewcommand{\arraystretch}{0.9}
\footnotesize
\begin{tabular}{@{}lcc@{}}
\toprule
& \multicolumn{2}{c}{WER (\%)} \\
\cmidrule(l){2-3}
Normalization& Utterance-level & Channel-level \\
\midrule
MN                & 13.55& 12.24\\
MVN               & 11.77& \textbf{11.47}\\
Sliding MVN       & 13.82& 11.95\\
\bottomrule
\end{tabular}
\vspace{-0.4cm}
\end{table}

Using the Hilbert transform at 25\,Hz cut-off as a fixed baseline, we evaluated the non-learnable normalization strategies in Table~\ref{tab:norm}. Channel-level normalization outperformed its utterance-level counterpart for every method, and MVN consistently improved over MN. The best result in this category, 11.47\% with channel-level MVN, suggests that frequency-specific gain and variance adjustments are essential for handling the variable energy distributions in children's speech. The sliding window was marginally worse than plain MVN at channel level (11.95\%) and substantially worse at utterance level (13.82\%), offering no benefit over full-utterance statistics. This could be due to the short, acoustically consistent MyST utterances leaving little within-utterance non-stationarity for local statistics to exploit.

\begin{table}[t]
\caption{Effect of filterbank type, frequency range, and normalization.
With Hilbert envelope extraction with $f_c{=}25$\,Hz.}
\label{tab:filterbank_pcen}
\centering
\setlength{\tabcolsep}{4pt}
\renewcommand{\arraystretch}{0.95}
\footnotesize
\begin{tabular}{@{}lcccc@{}}
\toprule
& \multicolumn{4}{c}{WER (\%)} \\
\cmidrule(l){2-5}
& \multicolumn{2}{c}{Chan.-level MVN} & \multicolumn{2}{c}{PCEN} \\
\cmidrule(lr){2-3} \cmidrule(l){4-5}
Filterbank & 0--8\,kHz & 150--8\,kHz & 0--8\,kHz & 150--8\,kHz \\
\midrule
Windowed-sinc & 11.47& 13.75& \textbf{11.08}& 11.62\\
Gabor         & 12.15& 11.76& 11.71& 11.71\\
\bottomrule
\end{tabular}
\end{table}

When comparing filter-bank architectures (Table~\ref{tab:filterbank_pcen}), the mel-spaced windowed-sinc filterbank over the full 0--8\,kHz range performed better than Gabor filter-bank under both channel-level MVN and PCEN normalization strategies. The mel-spaced windowed-sinc filterbank achieves a WER of 11.47\% with channel-level MVN (vs.\ 12.15\% for Gabor) and 11.08\% with PCEN (vs.\ 11.71\%). Narrowing the range to 150--8\,kHz, based on our pitch tracking analysis, degraded performance for the windowed-sinc filters (from 11.47\% to 13.75\% under MVN), while leaving the Gabor filters largely unchanged. This suggests that low-frequency components, while below the typical fundamental frequency range of child speakers, may still carry information useful to the downstream model. Finally, replacing non-learnable channel-level MVN with a learnable PCEN yielded our best configuration overall. The best configuration, a mel-spaced windowed-sinc filterbank (0--8\,kHz) with Hilbert envelope, 25\,Hz low-pass cutoff, and learnable PCEN, achieves a WER of \textbf{11.08\%}, a relative improvement of 15.8\% over the Whisper baseline. A paired bootstrap significance test \cite{Confidence_Intervals} with 1000 resamples, stratified by speaker confirmed significance at $p < 0.05$.

\begin{table}[t]
\caption{Comparison of our proposed system with the domain-adapted Kid-Whisper small multilingual~\cite{attia2024kidwhisperbridgingperformancegap} checkpoint on our MyST test split}
\label{tab:kidwhisper}
\centering
\footnotesize
\setlength{\tabcolsep}{5pt}
\renewcommand{\arraystretch}{1.05}
\begin{tabular}{@{}lcccc@{}}
\toprule
System & Sub. & Del. & Ins. & WER (\%) \\
\midrule
\makecell[l]{Kid-Whisper\\\footnotesize(fixed log-mel spectrogram)}
  & 15{,}786 & \textbf{3{,}984} & 9{,}732 & 17.42 \\
\makecell[l]{Ours\\\footnotesize(learnable frontend)}
  & \textbf{10{,}096} & 4{,}451 & \textbf{4{,}215} & \textbf{11.08} \\
\bottomrule
\end{tabular}
\vspace{-0.6cm}
\end{table}

\begin{table*}[!t]
\caption{Qualitative comparison of Kid-Whisper and proposed-frontend transcriptions on selected utterances illustrating each error type. Loops are abbreviated with repetition counts.}
\label{tab:erroranalysis}
\centering
\footnotesize
\renewcommand{\arraystretch}{1.0}
\newcolumntype{Y}{>{\ttfamily\arraybackslash}X}
\begin{tabularx}{\textwidth}{@{}lYYY@{}}
\toprule
Error type & \textrm{Reference} & \textrm{Kid-Whisper (fixed log-mel spectrogram)} & \textrm{Ours (learnable frontend)} \\
\midrule
Insertion &
it pumps blood &
\underline{that} it pumps blood &
it pumps blood \\
Insertion (loop) &
its the can can possibly hold &
that is the ["the"~$\times$148] &
its the can can possibly hold \\
Substitution &
the opposite poles &
the opposite \underline{pulls} &
the opposite poles \\
Deletion &
it provides the electricity pretty much &
it provides the electricity pretty much &
it provides electricity pretty much \\
\bottomrule
\end{tabularx}
\vspace{-0.4cm}
\end{table*}

\begin{figure}[t]
    \centering
    \includegraphics[width=\columnwidth]{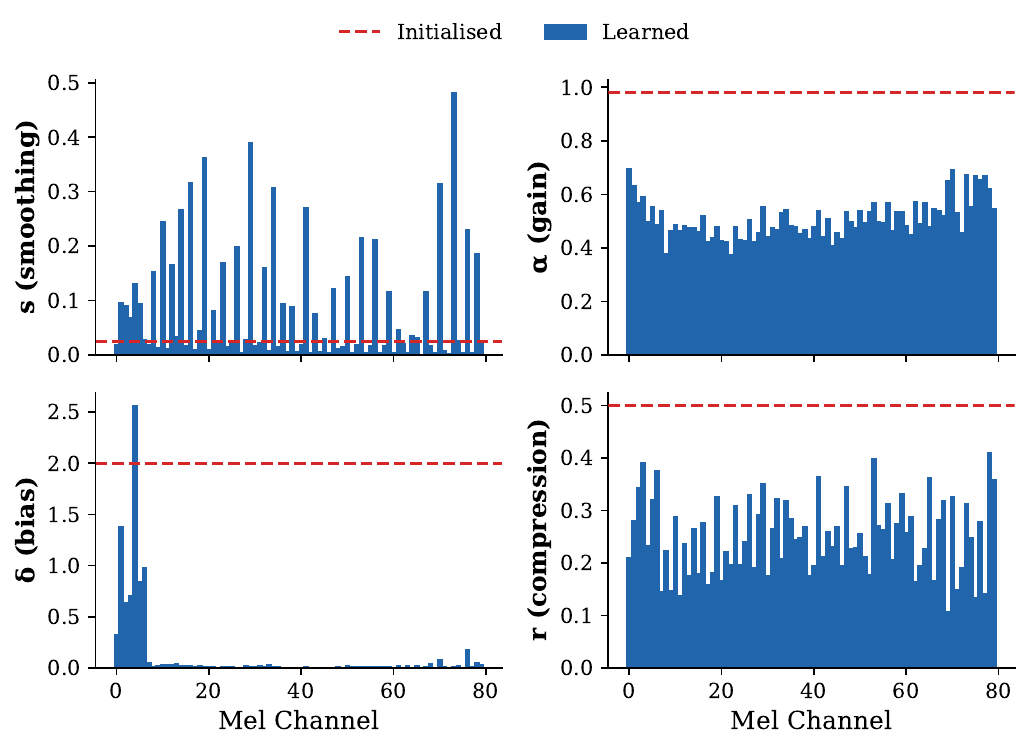}
    \caption{Learned per-channel PCEN parameters after joint optimization with the Whisper backend.}
    \label{fig:pcen_params}
    \vspace{-0.5cm}
\end{figure}
To better understand how the model adapts to children's speech, we also examine the learned PCEN parameters (Figure~\ref{fig:pcen_params}). All four parameters diverge from their initialization, revealing a channel-specific normalization strategy. While $\alpha$, $\delta$ and $r$ show interpretable frequency-dependent patterns, the smoothing coefficient $s$ varies erratically, suggesting multiple equally effective configurations exist across channels. The gain exponent $\alpha$ follows a U-shaped profile, applying stronger normalization at the spectral extremes likely stabilizing variable noise energy in the lowest channels and transient consonant bursts in the highest, while the mid-frequency formant bands require less aggressive gain control. The bias $\delta$ and compression $r$ act complementarily in the lowest bands such that high bias boosts weak sub-F0 content, which is then heavily compressed, forming a so-called \emph{boost-then-compress} strategy that extracts useful information without dominating the representation. These results suggest that temporal envelope with per-channel adaptive normalization is central to the observed improvement. The learned normalization captures spectral characteristics of children's speech that the conventional pipeline cannot.

Finally, we compare our best configuration against Kid-Whisper~\cite{attia2024kidwhisperbridgingperformancegap}, the released \textbf{multilingual} Whisper-small checkpoint fine-tuned on MyST with the standard log-mel frontend (Table~\ref{tab:kidwhisper}). To ensure a controlled comparison, we do not rely on published figures. Instead, we evaluate the released checkpoint ourselves on our test split, under the same text normalization. Both systems thus share the same backend, corpus, test data, and scoring, differing only in the frontend. On our split, Kid-Whisper yields 17.42\% against our 11.08\%, a 36.4\% relative improvement. Beyond this gap, the two approaches differ in what they adapt. Kid-Whisper improves children's ASR through data preprocessing and backend fine-tuning, leaving the log-mel frontend unchanged. Our system instead adapts the input representation as well, jointly optimizing the envelope frontend and the Whisper backend end-to-end. The frontend's gains therefore build on model-level adaptation rather than replacing it. This indicates that representation-level and model-level adaptation are complementary.

Figure~\ref{fig:spectrogram} compares the log-mel features with that from our proposed frontend for the same utterance. Relative to the log-mel spectrogram in (a), the envelope-based representation in (b) exhibits cleaner non-speech regions and sharper temporal boundaries at speech onsets and offsets, while preserving the harmonic structure in the lower-to-mid mel bands where much of the energy in children's voiced speech resides. These properties provide the encoder with a more robust representation of children's speech, which we suggest contributes to the observed WER improvement.

Inspecting the per-utterance differences shows that the improvement is generally broad rather than concentrated. Most of the gains come from many small reductions across the test set rather than a few large ones, indicating that the frontend benefits children's speech generally. Relative to Kid-Whisper (fixed log-mel spectrogram), the proposed learnable frontend reduces insertions by 57\% (9,732 to 4,215) and substitutions by 36\% (15,786 to 10,096). Table~\ref{tab:erroranalysis} illustrates each type of error. The reduction in substitutions broadly reflects more accurate word recognition, as in \textit{the opposite poles}, which the baseline renders as \textit{the opposite pulls}. The reduction in insertions reflects suppressed over-generation, ranging from spurious words to extreme cases where the baseline collapses into repetitive loops on short or low-energy input. Although such loops are not frequent, they illustrate a degenerate behavior the frontend avoids. This is consistent with the stabilising role of PCEN identified in Figure~\ref{fig:pcen_params}. The per-channel gain control on low-energy frames prevents the near-zero embeddings that trigger collapse, making the representation more robust on short, low-energy input. The frontend does, however, introduce occasional deletions as in \textit{it provides the electricity} rendered as \textit{it provides electricity}, accounting for the increase in deletion errors.

\section{Conclusion}
\vspace{-0.1cm}
We presented a temporal-envelope frontend for Whisper-based children's ASR showing that adapting the acoustic representation complements backend adaptation to improve recognition of children's speech. We summarize our main contributions and findings as follows:
\begin{itemize}
    \item We \emph{adapt} the acoustic frontend, unlike most Whisper-based children's ASR work that keeps the log-mel frontend \emph{fixed}. The proposed pipeline uses mel-spaced time-domain filtering, Hilbert envelope extraction, temporal smoothing, RMS pooling, and learnable PCEN.
    \item We evaluate temporal-envelope features for children's ASR with Whisper, whereas prior envelope-based work targeted adult ASR, robustness, or keyword spotting.
    \item Systematic MyST ablations identify full-band $0$--$8$ kHz windowed-sinc filters, Hilbert envelopes, a $25$ Hz cutoff, and learnable PCEN as the best configuration.
    \item The proposed frontend reduces WER from $13.16\%$ to $11.08\%$ over the identically fine-tuned Whisper baseline ($15.8\%$ relative reduction), and improves over the released Kid-Whisper multilingual checkpoint ($17.42\%$) on the same cleaned test split.
    \item Error analysis shows reduced substitutions and insertions, including repetitive decoding loops, though deletions increase modestly.
\end{itemize}
These results suggest that temporal-envelope representations and learnable per-channel normalization are effective complements to existing children's ASR adaptation methods. Future work will evaluate additional corpora, larger ASR models, and causal filtering for streaming recognition.



\bibliographystyle{IEEEtran}
\bibliography{references}

\end{document}